# Free-electron Brewster radiation

Ruoxi Chen[1,2,#], Jialin Chen[1,2,3,#], Zheng Gong[1,2], Xinyan Zhang[1,2], Xingjian Zhu[4], Yi Yang[5], Ido Kaminer[3,*], Hongsheng Chen[1,2,6,7,*], Baile Zhang[8,9], and Xiao Lin[1,2,*]

[1]*Interdisciplinary Center for Quantum Information, State Key Laboratory of Extreme Photonics and Instrumentation, ZJU-Hangzhou Global Scientific and Technological Innovation Center, College of Information Science & Electronic Engineering, Zhejiang University, Hangzhou 310027, China.*

[2]*International Joint Innovation Center, the Electromagnetics Academy at Zhejiang University, Zhejiang University, Haining 314400, China.*

[3]*Department of Electrical and Computer Engineering, Technion-Israel Institute of Technology, Haifa 32000, Israel.*

[4]*School of Physics, Zhejiang University, Hangzhou 310027, China.*

[5]*Department of Physics, University of Hong Kong, Hong Kong 999077, China.*

[6]*Key Laboratory of Advanced Micro/Nano Electronic Devices & Smart Systems of Zhejiang, Jinhua Institute of Zhejiang University, Zhejiang University, Jinhua 321099, China.*

[7]*Shaoxing Institute of Zhejiang University, Zhejiang University, Shaoxing 312000, China.*

[8]*Division of Physics and Applied Physics, School of Physical and Mathematical Sciences, Nanyang Technological University, Singapore 637371, Singapore.*

[9]*Centre for Disruptive Photonic Technologies, Nanyang Technological University, Singapore 637371, Singapore.*

[#]*These authors contribute equally.*

*Corresponding authors: xiaolinzju@zju.edu.cn (X. Lin); kaminer@technion.ac.il (I. Kaminer); hansomchen@zju.edu.cn (H. Chen)

**Free-electron radiation offers an enticing route to create light emission at arbitrary spectral regime. However, this type of light emission is generally weak, which is intrinsically limited by the weak particle-matter interaction and unavoidably impedes the development of many promising applications, such as the miniaturization of free-electron radiation sources and high-energy particle detectors. Here we reveal a mechanism to enhance the particle-matter interaction by exploiting the pseudo-Brewster effect of gain materials – presenting an enhancement of at least four orders of magnitude for the light emission. This mechanism is enabled by the emergence of *an unprecedented phase diagram* that maps all phenomena of free-electron radiation into three distinct phases in a gain-thickness parameter space, namely the conventional, intermediate, and Brewster phases, when an electron penetrates a dielectric slab with a modest gain and a finite thickness. Essentially, our revealed mechanism corresponds to the free-electron radiation in the Brewster phase, which also uniquely features ultrahigh directionality,**

**always at the Brewster angle, regardless of the electron velocity. Counterintuitively, we find that the intensity of this free-electron Brewster radiation is insensitive to the Fabry-Pérot resonance condition and thus the variation of slab thickness, and moreover, a *weaker* gain could lead to a *stronger* enhancement for the light emission. The scheme of free-electron Brewster radiation, especially along with its compatibility with low-energy electrons, may enable the development of high-directionality high-intensity light sources at any frequency.**

Free-electron radiation [1-8] is a fundamental process of light emission that originates from the interaction between fast-moving charged particles (e.g., a swift electron or ion) and optical matter. One powerful capability of free-electron radiation is that it can create light emission at any frequency, ranging from the microwave to X-ray regimes. The light emission could be exploited to not only implement novel types of radiation *sources* (e.g. free-electron lasers [9-12] and high-power microwave sources [13,14]), but also be measured to detect the information of *particles* (e.g. high-energy particle detector [15-19] and particle-beam diagnosis [20,21]) or optical *matter* (e.g. electron microscopy [22] and medical imaging [23]). Therefore, free-electron radiation is of paramount importance to many areas of science, as varied as high-energy physics, astronomy, cosmology, nanophotonics, plasmonics, material science, and bio-medicine, and is currently opening new frontiers across these areas.

However, the intensity of free-electron radiation is generally weak due to the weak interaction between each particle and optical matter. To enable many of the applications envisioned in this field [18,24-31], free-electron radiation should be enhanced. There are currently three main ways to achieve such an enhancement. One way is to increase the current density of particle beams, such as the high-current electron beam widely used in the high-power microwave source [13,14]. The second way is to elongate the interaction length. The third way is to accelerate the particle beam to almost the speed $c$ of light in free space. For example, the first free-electron laser [32], as developed by Madey in 1971, uses a 43 mega-electron-volt (MeV) electron beam and a five-meter-long wiggler to amplify the light emission. Another example is the Cherenkov detector [15-19] – a famous particle detector based on Cherenkov radiation [33,34] – whose radiator generally has a meter-scale length and is designed particularly for the identification of high-energy particles with their kinetic energy up to the giga-eV (GeV) scale.

Nevertheless, at the base of all these is still a weak particle-matter interaction. The weak interaction severely impedes the development of many more enticing applications of free-electron radiation, such as the miniaturization of free-electron radiation sources [22,28-30,35,36] and high-energy particle detectors [18,19,37]. The realization of such applications could boost the on-chip integration of free-electron light sources (e.g. in the THz/X-ray regimes) and to facilitate the direct detection of high-energy particles in outer space. Despite the long research history of free-electron radiation [38-49], there remains a need for fundamentally different mechanisms to enhance the particle-matter interaction, especially for low-energy particles.

Here we propose a mechanism to enhance the particle-matter interaction by exploiting the pseudo-Brewster effect [50-57] of gain materials, which could lead to the light emission with an enhancement of many (e.g. four) orders of magnitude. While the optical gain provides a universal way to amplify the light emission for most optical systems, the influence of optical gain on the free-electron radiation (including its intensity and directionality) is underexplored; especially, the pseudo-Brewster effect of gain materials has never been connected to the particle-matter interaction. We find that the connection of pseudo-Brewster effect to the particle-matter interaction enables us to map all phenomena of free-electron radiation into an unprecedented phase diagram in a gain-thickness parameter space, when an electron perpendicularly penetrates a dielectric slab with a modest gain and a finite gain. We highlight that our proposed mechanism is not purely due to the existence of optical gain but enabled by the emergence of this exotic phase diagram of free-electron radiation, which could be categorized into three distinct phases, namely the conventional, intermediate, and Brewster phases. Essentially, our proposed mechanism corresponds to the free-electron radiation in the Brewster phase. Counterintuitively, the revealed free-electron radiation in the Brewster phase has three exotic features. First, we find that the optical gain in the Brewster phase can not only substantially enhance the intensity of light emission but also its directionality. As a result, the free-electron radiation in the Brewster phase is uniquely featured with ultrahigh directionality, always at the Brewster angle, irrespective to the electron velocity. Second, we find the optical gain in the Brewster phase can disable the Fabry-Pérot resonance condition. Correspondingly, the intensity of free-electron radiation in the Brewster phase becomes in-sensitive to the variation of slab thickness. Third, we find that a weaker optical gain can lead to a stronger enhancement of free-electron radiation in the Brewster phase, as fundamentally governed by the pseudo-Brewster effect of gain materials.

To place our results in the proper context, Cherenkov radiation [33,34] – a famous type of free-electron radiation inside a homogeneous matter – is created only when the electron moves with a velocity $v$ larger than the phase velocity $c/n$ of light in that matter (known as the Cherenkov threshold $v_\text{th}$), namely $v \geq v_\text{th} = c/n$, where $n$ is the refractive index of matter. Although Cherenkov radiation is highly directional, its radiation angle $\theta$ – known as the Cherenkov angle – is sensitive to the electron velocity, as governed by the Frank-Tamm formula of $cos\theta = c/nv$ [34,58]. Transition radiation [18,44,59-62] – another typical type of free-electron radiation – could occur when the electron moves across an inhomogeneous region (e.g. an optical interface) at any speed. Still, transition radiation is generally suffered from its low directionality and low intensity. Transition radiation becomes relatively directional, with its maximum appearing at the radiation angle of $\theta = 1/\gamma$, only when the electron is highly relativistic (e.g. when $\gamma > 10$) [63], where $\gamma = 1/\sqrt{1 - v^2/c^2}$ is the Lorentz factor. Remarkably, our revealed free-electron radiation in the Brewster phase has its directionality property (e.g. the dependence of radiation directionality on the electron velocity) completely different from that of Cherenkov radiation, transition radiation and other types of free-electron radiation [6-8,22,28-30,62], such as Smith-Purcell radiation, synchrotron radiation, and bremsstrahlung radiation; therefore, it corresponds to a brand-new form of free-electron radiation. Because of its unique directionality, we suggest to denote the free-electron radiation in the Brewster phase as *free-electron Brewster radiation*. Due to the simultaneous enhancement of its intensity and directionality, the free-electron Brewster radiation may offer a feasible route for the development of novel light sources based on ultralow-energy electrons, with their kinetic energies down to the scale of several electron volts (eV).

We begin with the conceptual demonstration of free-electron Brewster radiation in Fig. 1a-e. We consider a gain material (namely region 2 in Fig. 1a) with a relative permittivity of $\varepsilon_{r,2} = 2 - 0.1i$ at the working wavelength of $\lambda_0$ in free space; this gain in practice can be implemented, for example, by using negative-resistance components [64-67] (e.g. microwave tunnel diodes) and optically pumped dye molecules [68,69] (e.g. Rhodamine 800 dye molecules). A moving electron with $\bar{v} = \hat{z}0.5c$ is adopted, and both the slab's upper region (region 1) and lower region (region 3) are free space. This way, there is no Cherenkov radiation in the proposed system in Fig. 1, since the electron velocity is below the Cherenkov threshold. The slab has a finite thickness of $d = 50\lambda_0$; to avoid the potential scattering of swift electrons, a hole with its center along the electron trajectory could be drilled for the slab [Fig. S2a], which is feasible especially for microwave gain slabs [64-67].

When the free-electron Brewster radiation occurs in Fig. 1a, strong plane-like waves propagate into the lower region. As shown by the forward angular spectral energy density $U_F(\theta_F)$ in Fig. 1c-d, the free-electron Brewster radiation is highly directional, and its radiation peak shows up at $\theta_{F,peak} = 54.7°$, where $\theta_F$ represents the forward radiation angle. Actually, we have $\theta_{F,peak} = \theta_{Brew,pseudo}$, where $\theta_{Brew,pseudo} = Re\left(\arctan(\sqrt{\varepsilon_{r,2}/\varepsilon_{r,3}})\right)$ is the pseudo-Brewster angle at the interface between the gain slab and its lower region with a relative permittivity of $\varepsilon_{r,3} = 1$. This directional feature of free-electron Brewster radiation from a gain slab in Fig. 1a is completely different from that of conventional free-electron radiation from a transparent or lossy slab. For example, if the gain slab is replaced by a transparent slab with $\varepsilon_{r,2} = 2$, there are only spherical-like waves in the lower region in Fig. 1b. The forward angular spectral energy density in Fig. 1c&e further confirms that the conventional free-electron radiation is of low directionality and low intensity. As a result, we find in Fig. 1c that the intensity of free-electron Brewster radiation is four orders of magnitude larger than that of conventional free-electron radiation at the radiation angle of $\theta_F = \theta_{Brew,pseudo}$. That is, the gain slab behaves as a powerful angular amplifier, which is capable to amplify the free-electron radiation predominantly at the pseudo-Brewster angle.

We now proceed to explain the occurrence of free-electron Brewster radiation in Fig. 2. By following Ginzburg and Frank's theory of free-electron radiation [18,44,59-61,70], the forward radiation field in the lower region can be written as $E^R_{z,forward}(\bar{r}, t) = \iiint d\omega d\bar{k}_\perp E^R_{z,forward}(z) e^{i(\bar{k}_\perp \cdot \bar{r}_\perp - \omega t)}$, where

$$E^R_{z,forward}(z) = A_{forward} e^{ik_{z,3}(z-d)} \tag{1}$$

$\bar{k}_\perp = \hat{x}k_x + \hat{y}k_y$, $k_\perp = |\bar{k}_\perp|$, $\bar{r}_\perp = \hat{x}x + \hat{y}y$, $k_{z,j}$ is the $z$ component of wavevector in region $j$ ($j = 1$, 2 or 3), $k_{z,j}^2 + k_\perp^2 = k_j^2$, and $k_j^2 = \varepsilon_{r,j}\omega^2/c^2$. After some calculation (see supplementary sections S1-S2), the forward radiation coefficient $A_{forward}$ is formulated as

$$A_{forward} = a^+_{2,3} + a^+_{1,2} \frac{e^{ik_{z,2}d}}{1 - R_{2,1}R_{2,3}e^{2ik_{z,2}d}} + a^-_{2,3} \frac{R_{2,1}e^{2ik_{z,2}d}}{1 - R_{2,1}R_{2,3}e^{2ik_{z,2}d}} \tag{2}$$

where $a^\pm_{j,j+1}$ and $R_{j,j+1} = -R_{j+1,j}$ are related to the radiation coefficient and the reflection coefficient at the interface between region $j$ and region $j + 1$, respectively.

When region 2 is filled with a gain material, we have $Im(k_{z,2}) < 0$. Correspondingly, we have $|e^{2ik_{z,2}d}| \to \infty$ and $|R_{2,1}R_{2,3}e^{2ik_{z,2}d}| \gg 1$, if $d/\lambda_0$ is large enough. This way, we always have $\lim_{d\to\infty} \frac{e^{ik_{z,2}d}}{1-R_{2,1}R_{2,3}e^{2ik_{z,2}d}} = 0$ and $\lim_{d\to\infty} \frac{R_{2,1}e^{2ik_{z,2}d}}{1-R_{2,1}R_{2,3}e^{2ik_{z,2}d}} = \frac{-1}{R_{2,3}}$ in equation (2). Under this scenario, equation (2) is simplified to

$$\lim_{d\to\infty} A_{\text{forward}} = a_{2,3}^+ - a_{2,3}^- \frac{1}{R_{2,3}} \tag{3}$$

In equation (3), $a_{2,3}^+$ and $a_{2,3}^-$ are generally in the same order of magnitude [as shown in Fig. S3], and we could have $\left| a_{2,3}^- \frac{1}{R_{2,3}} \right| \gg |a_{2,3}^+|$ if $|R_{2,3}| \to 0$. For gain materials, $|R_{2,3}| \to 0$, instead of $|R_{2,3}| = 0$, is achievable at the pseudo-Brewster angle of $\theta_{\text{Brew,pseudo}}$, according to the pseudo-Brewster effect of gain or lossy materials [50-57]. Moreover, if $Im(\varepsilon_{r,2})$ is reasonably large as shown in Fig. S5c, $\theta_{\text{Brew,pseudo}}$ is insensitive to $Im(\varepsilon_{r,2})$, and we have $\theta_{\text{Brew,pseudo}} = \theta_{\text{Brew}}$, where $\theta_{\text{Brew}} = \arctan(\sqrt{Re(\varepsilon_{r,2})/\varepsilon_{r,3}})$ is the Brewster angle according to the Brewster effect of transparent materials [58,71-73].

With this knowledge of the pseudo-Brewster effect of gain materials, we conclude that the term of $a_{2,3}^- \frac{1}{R_{2,3}}$ plays a determinant role in equation (3), while the contribution from $a_{2,3}^+$ is negligible, if the radiation angle is close to the Brewster angle (namely $\theta_F \to \theta_{\text{Brew}}$). Then equation (3) could be further reduced to

$$\lim_{\substack{d\to\infty \\ \theta_F \to \theta_{\text{Brew}}}} A_{\text{forward}} = -a_{2,3}^- \frac{1}{R_{2,3}} \tag{4}$$

Equation (4) indicates that the maximum of $|A_{\text{forward}}|$, along with the minimum value of $|R_{2,3}|$, would appear at the Brewster angle. In other words, the free-electron Brewster radiation with an ultrahigh directionality and an enhanced intensity would always occur at the Brewster angle as long as the slab thickness is large enough; see the schematic of free-electron Brewster radiation in *k* space in Fig. 2a.

Upon close inspection of equation (2), the pole of $|A_{\text{forward}}|$ directly corresponds to the pole of $\frac{R_{2,1}e^{2ik_{z,2}d}}{1-R_{2,1}R_{2,3}e^{2ik_{z,2}d}}$, a key factor that originates from the wave resonance inside the slab and could manifest the eigenmodes (e.g. guidance modes, leaky modes) supported by the slab. To facilitate the understanding of free-electron Brewster radiation, the poles of $|A_{\text{forward}}|$ from a sufficiently-thick gain slab is plotted in the complex $k_\perp$ plane in Fig. 2c. There are three distinct types of poles in Fig. 2c. The first type of poles has $Re(k_\perp) = k_{\perp,\text{guidance}}$, where $k_{\text{air}} < k_{\perp,\text{guidance}} < Re(k_2)$ and $k_{\text{air}} = \omega/c$. As such, all these poles correspond to the guidance modes, which would not contribute to the light emission into the far field. The other two types of poles have $Re(k_\perp) < k_{\text{air}}$, and hence, their corresponding eigenmodes are intrinsically leaky and can couple into free space. To be specific, the second type of poles corresponds to the conventional leaky modes, whose behavior to some extents are related to the Fabry-Pérot resonance condition of $arg(R_{2,1} R_{2,3} e^{2ik_{z,2}d}) = 2\pi$, since their pole position (e.g. their value of $Re(k_\perp)$) is sensitive to the variation of slab thickness; see the influence of slab thickness in Figs. S6-S8. By contrast, the third type only has one pole in Fig. 2c. Remarkably, its position becomes irrelevant to the Fabry-Pérot resonance condition and appears always at $Re(k_\perp) = k_{\perp,\text{Brew}}$ for a sufficiently-thick slab, where $k_{\perp,\text{Brew}} = k_{\text{air}} \sin\theta_{\text{Brew}}$. Due to its intrinsic connection with the pseudo-Brewster effect of gain materials, this special eigenmode is termed as the Brewster leaky mode in Fig. 2c. The Brewster leaky mode once excited would have a larger contribution to the far-field free-electron radiation than the conventional leaky modes, since the intensity of forward free-electron radiation is proportional to the value of $|A_{\text{forward}}|^2$ at the real-$k_\perp$ axis and the pole position of the third type in the complex $k_\perp$ plane in Fig. 2c is much closer to the real-$k_\perp$ axis than that of the second type; see detailed discussions in Figs. S6-S8. Therefore, the occurrence of free-electron Brewster radiation is attributed to the excitation of Brewster leaky modes, in accordance to our above analysis for equation (4). As background, Fig. 2d shows that $|A_{\text{forward}}|$ from a transparent slab only has the first type of poles.

From the above analyses for equations (2-4), the slab thickness $d$ should be sufficiently large to enable the appearance of free-electron Brewster radiation. We show in Fig. 3a the influence of $d$ on the forward free-electron radiation. There are mainly two well-separated groups of radiation peaks in Fig. 3a, if $d$ is relatively small (e.g. $d < d_c = 11\lambda_0$). These peaks tend to merge to the Brewster angle if $d$ increases, and then the free-electron Brewster radiation appears if $d$ is relatively large (e.g. $d > d_{\text{Brew}} = 20\lambda_0$ in Fig. 3a). To quantify this tendency, the angular deviation between these two groups of radiation peaks is defined as $\Delta\theta = \theta_{\text{max,right}} - \theta_{\text{max,left}}$, where $\theta_{\text{max,left}}$ and $\theta_{\text{max,right}}$ correspond to the angular positions of radiation peaks within the range of $\theta_F \in [0°, \theta_{\text{Brew}}]$ and $\theta_F \in [\theta_{\text{Brew}}, 90°]$, respectively. Fig. 3b shows that the value of $\Delta\theta$ oscillates randomly and $\Delta\theta > \Delta\theta_c$, if $d < d_c$, where $\Delta\theta_c = 10°$. The value of $\Delta\theta$ becomes to decrease with the increase of $d$ if $d > d_c$; moreover, we have $\Delta\theta_{\text{Brew}} < \Delta\theta < \Delta\theta_c$ if $d_c < d < d_{\text{Brew}}$ and $\Delta\theta < \Delta\theta_{\text{Brew}}$ if $d > d_{\text{Brew}}$, where $\Delta\theta_{\text{Brew}} = 0.5°$.

We then denote the free-electron Brewster radiation with $\Delta\theta < \Delta\theta_{\text{Brew}}$ as the Brewster phase of free-electron radiation, free-electron radiation with $\Delta\theta_{\text{Brew}} < \Delta\theta < \Delta\theta_c$ as the intermediate phase, and the conventional free-electron radiation with $\Delta\theta > \Delta\theta_c$ as the conventional phase, due to their distinct radiation properties in these three phases (e.g. directionality revealed in Fig. 3a-b and intensity in Fig. 3a&c).

Due to the exotic feature of forward free-electron radiation at the Brewster angle in Fig. 3a, we further investigate in Fig. 3c the radiation intensity at the Brewster angle, namely $U_F(\theta_{\text{Brew}})$, as a function of the slab thickness $d$. The value of $U_F(\theta_{\text{Brew}})$ in Fig. 3c tends to increase with $d$ if $d < d_{\text{Brew}}$; and it finally saturates to a constant value of $U_{F,\infty}$ and becomes insensitive to the variation of $d$ if $d > d_{\text{Brew}}$, where

$$U_{F,\infty} = \lim_{\substack{d \to \infty \\ \theta_F \to \theta_{\text{Brew}}}} U_F(\theta_{\text{Brew}}) \propto \lim_{d \to \infty} |A_{\text{forward}}|^2 \propto \lim_{\theta_F \to \theta_{\text{Brew}}} \left|\frac{1}{R_{2,3}}\right|^2 \qquad (5)$$

Equation (5) indicates that the intensity of free-electron Brewster radiation at the Brewster angle is determined merely by the optical response of gain materials, namely $-\text{Im}(\varepsilon_{r,2})$, as shown in Fig. 3c-d. Counterintuitively, the value of $U_{F,\infty}$ increases while $|\text{Im}(\varepsilon_{r,2})|$ decreases in Fig. 3d. That is, the intensity of free-electron Brewster radiation can be further enhanced by using a slab with a weaker gain; for example, this enhancement reaches over ten times if $|\text{Im}(\varepsilon_{r,2})|$ decreases from 0.1 to 0.025 in Fig. 3c-d.

From Fig. 3a-d, the free-electron radiation from a gain slab, especially its directionality, strongly correlates to the slab thickness and the optical gain. We characterize the directionality by $\Delta\theta$ in the thickness-gain parameter space in Fig. 3e and find clear boundaries between $\Delta\theta > \Delta\theta_c$, $\Delta\theta_{\text{Brew}} < \Delta\theta < \Delta\theta_c$ and $\Delta\theta < \Delta\theta_{\text{Brew}}$ in the thickness-gain space. Therefore, Fig. 3e indicates the existence of an exotic phase diagram for free-electron radiation, which is divided into three distinct phases, namely the conventional, intermediate, and Brewster phases.

We further show in Fig. 4a that the occurrence of free-electron Brewster radiation is robust to the variation of electron velocity, and actually, it could occur at any electron velocity. On the one hand, the ultrahigh directionality of free-electron Brewster radiation from a gain slab shows up always at the sole Brewster angle in Fig. 4a, which is independent on the electron velocity and distinctly different from the low directionality of conventional free-electron radiation from a transparent slab in Fig. 4b. On the other hand, Fig. 4c shows that the intensity of free-electron Brewster radiation is at least four orders of magnitude larger than that of conventional free-electron radiation. This large enhancement also occurs at any electron velocity, even when the electron velocity is far below the light speed (e.g. $v/c < 10^{-3}$ in Fig. 4c).

In conclusion, we have revealed the emergence of an unprecedented phase diagram in the gain-thickness parameter space – which could map all free-electron radiation phenomena into three distinct phases, namely the conventional, intermediate, and Brewster phases – by connecting the pseudo-Brewster effect of gain materials to the particle-matter interaction. If the free-electron radiation is in the Brewster phase, the free-electron Brewster radiation appears and is uniquely featured with an ultrahigh directionality, which always occurs at the Brewster angle and is robust to the electron velocity due to the pseudo-Brewster effect of gain materials. Therefore, the revealed free-electron Brewster radiation represents a new form of free-electron radiation and has its directionality property completely different from all other types of free-electron radiation, including Cherenkov radiation and transition radiation, whose directionality are sensitive to the electron velocity. Moreover, we find the pseudo-Brewster effect could substantially enhance the intensity of free-electron Brewster radiation by many orders of magnitude. Counterintuitively, we further find that the intensity of free-electron Brewster radiation is insensitive to the Fabry-Pérot resonance condition and the variation of the slab thickness; moreover, a weaker gain could lead to the free-electron Brewster radiation with a higher intensity. Our work indicates that the free-electron Brewster radiation is promising for the development of advanced light sources, since it could offer an enticing route to improve both the intensity and directionality of light emission, which is feasible for any particle velocity, even for ultralow-energy electrons with their kinetic energy down to the eV scale, and is in principle feasible for any frequency, ranging from the microwave to X-ray regimes.

**Acknowledgement**
X.L. acknowledges the support partly from the National Natural Science Fund for Excellent Young Scientists Fund Program (Overseas) of China, the National Natural Science Foundation of China (NSFC) under Grant No. 62175212, Zhejiang Provincial Natural Science Fund Key Project under Grant No. LZ23F050003, the Fundamental Research Funds for the Central Universities (2021FZZX001-19), and Zhejiang University Global Partnership Fund. H.C. acknowledges the support from the Key Research and Development Program of the Ministry of Science and Technology under Grants No. 2022YFA1404704, 2022YFA1404902, and 2022YFA1405200, the National Natural Science Foundation of China (NNSFC) under Grants No.11961141010 and No. 61975176. J.C. acknowledges the support from the Chinese Scholarship Council (CSC No. 202206320287). Y.Y. acknowledges the support from the start-up fund of the University of Hong Kong and the National Natural Science Foundation of China Excellent Young Scientists Fund (HKU 12222417). I.K. acknowledges the support from the Israel Science Foundation under Grant No. 3334/19 and the Israel Science Foundation under Grant No. 830/19. B.Z. acknowledges the support from Singapore National Research Foundation Competitive Research Program No. NRF-CRP23-2019-0007.

**Author contributions**
R.C. and X.L. initiated the idea; R.C. and J.C. performed the calculation; Z.G., X.Z., X.Z., Y.Y., I.K., H.C., and B.Z. helped to analyze data and interpret detailed results; R.C., X.L., and I.K. wrote the manuscript; X.L., I.K., B.Z., and H.C. supervised the project.

## Competing interests
The authors declare no competing interests.

## Methods

**Derivation of free-electron radiation from an interface.** Free-electron radiation from a single interface between region $j$ and region $j+1$ (see the structural schematic in Fig. S1) can be rigorously calculated by following Ginzburg and Frank's theory of free-electron radiation within the framework of classical electrodynamics. Detailed derivation is provided in supplementary section S1. If the interface between region $j$ and region $j+1$ is at the plane of $z=0$, the backward radiation coefficient can be obtained as $a_{j,j+1}^{0,-} = \frac{\frac{v}{c}\frac{k_\perp^2 c^2}{\omega^2 \varepsilon_{r,j}}(\varepsilon_{r,j+1}-\varepsilon_{r,j})\left(1-\frac{v^2}{c^2}\varepsilon_{r,j}+\frac{v}{c}\frac{k_{z,j+1}}{\omega/c}\right)}{\left(1-\frac{v^2}{c^2}\varepsilon_{r,j}+\frac{k_\perp^2 v^2}{\omega^2}\right)\left(1+\frac{v}{c}\frac{k_{z,j+1}}{\omega/c}\right)\left[\varepsilon_{r,j}\frac{k_{z,j+1}}{\omega/c}+\varepsilon_{r,j+1}\frac{k_{z,j}}{\omega/c}\right]}$, and similarly, the forward radiation coefficient can be expressed as $a_{j,j+1}^{0,+} = \frac{\frac{v}{c}\frac{k_\perp^2 c^2}{\omega^2 \varepsilon_{r,j+1}}(\varepsilon_{r,j+1}-\varepsilon_{r,j})\left(1-\frac{v^2}{c^2}\varepsilon_{r,j+1}-\frac{v}{c}\frac{k_{z,j}}{\omega/c}\right)}{\left(1-\frac{v^2}{c^2}\varepsilon_{r,j+1}+\frac{k_\perp^2 v^2}{\omega^2}\right)\left(1-\frac{v}{c}\frac{k_{z,j}}{\omega/c}\right)\left[\varepsilon_{r,j}\frac{k_{z,j+1}}{\omega/c}+\varepsilon_{r,j+1}\frac{k_{z,j}}{\omega/c}\right]}$, where the superscript '+' and '−' correspond to the forward and backward radiation, respectively.

**Derivation of free-electron radiation from a slab.** When a fast electron perpendicularly penetrates through a slab (see the structural schematic in Fig. S2), the related free-electron radiation can also be rigorously calculated by following Ginzburg and Frank's theory of free-electron radiation. Based on the result of free-electron radiation from an interface, the forward radiation coefficient $A_{\text{forward}}$ can be expressed as $A_{\text{forward}} = a_{2,3}^+ + a_{1,2}^+ \frac{e^{ik_{z,2}d}}{1-R_{2,1}R_{2,3}e^{2ik_{z,2}d}} + a_{2,3}^- \frac{R_{2,1}e^{2ik_{z,2}d}}{1-R_{2,1}R_{2,3}e^{2ik_{z,2}d}}$, where $a_{2,3}^+ = a_{2,3}^{0,+} e^{i\frac{\omega}{v}d}$, $a_{1,2}^+ = a_{1,2}^{0,+} T_{2,3}$, and $a_{2,3}^- = a_{2,3}^{0,-} T_{2,3} e^{i\frac{\omega}{v}d}$ and $T_{2,3}$ is the transmission coefficient at the interface between region 2 and 3. Detailed derivation is provided in supplementary section S2.

**Comparison between $|a_{2,3}^+|$ and $|a_{2,3}^-|$.** We show in supplementary section S3 and Fig. S3 that the values of $|a_{2,3}^+|$ and $|a_{2,3}^-|$ are in the same order of magnitude.

**Angular spectral energy density $U_F(\theta_F)$ of forward free-electron radiation at different electron velocities.** In supplementary section S4, the detailed calculation of angular spectral energy density $U_F(\theta_F, \omega)$ of forward free-electron radiation is provided. Based on the results of radiation field, the forward angular spectral energy density can be obtained as $U_F(\theta_F) = \frac{\varepsilon_{r,3}^{3/2} q^2 \cos^2\theta_F}{4\pi^3 \varepsilon_0 c \sin^2 \theta_F} |A_{\text{forward}}|^2$. Without loss of generality, we discuss in supplementary section S5 the influence of the electron velocity on the forward angular spectral energy density for both gain and transparent slabs; see Fig. S4.

**Pseudo-Brewster effect.** In supplementary section S6, we briefly introduce the Brewster effect of transparent materials and the pseudo-Brewster effect of gain or lossy materials. We show in Fig. S5 that if $|Im(\varepsilon_{r,2})|$ is reasonably large, the pseudo-Brewster angle $\theta_{\text{Brew,pseudo}} = Re\left(\arctan\left(\sqrt{\varepsilon_{r,2}/\varepsilon_{r,3}}\right)\right)$ is generally insensitive to $|Im(\varepsilon_{r,2})|$, and we approximately have $\theta_{\text{Brew,pseudo}} = \theta_{\text{Brew}}$, where $\theta_{\text{Brew}} = $ $\arctan\left(\sqrt{Re(\varepsilon_{r,2})/\varepsilon_{r,3}}\right)$ is the Brewster angle for transparent system.

**Poles of forward radiation coefficient $A_{forward}$ in the complex $k_\perp$ plane.** To facilitate the understanding of free-electron Brewster radiation, we show the poles of forward radiation coefficient $A_{forward}$ in the complex $k_\perp$ plane in supplementary section S7. Moreover, the influence of slab thickness on the pole positions are discussed in Figs. S6-S8.

**Dependence of the angular deviation $\Delta\theta$ on the slab thickness $d$ at different optical gains.** Without loss of generality, we show in Fig. S9 the dependence of $\Delta\theta$ on $d$ at different optical gains. Detailed discussion is provided in supplementary section S8.

**Free-electron radiation from a lossy slab.** For comparison, we discuss the free-electron radiation from a lossy slab in the supplementary section S9. The influence of loss on the forward free-electron radiation is shown in Fig. S10.

## Data availability
The data that support the findings of this study are available from the corresponding authors upon reasonable request.

## Reference

[1] Fisher, A. et al. Single-pass high-efficiency terahertz free-electron laser. *Nat. Photon.* **16**, 441-447 (2022).
[2] Konečná, A., Iyikanat, F. & de Abajo, F. J. G. Entangling free electrons and optical excitations. *Sci. Adv.* **8**, eabo7853 (2022).
[3] Zhang, D. et al. Coherent surface plasmon polariton amplification via free-electron pumping. *Nature* **611**, 55-60 (2022).
[4] Sapra, N. V. et al. On-chip integrated laser-driven particle accelerator. *Science* **367**, 79-83 (2021).
[5] Bergmann, U. et al. Using X-ray free-electron lasers for spectroscopy of molecular catalysts and metalloenzymes. *Nat. Rev. Phys.* **3**, 264-282 (2021).
[6] Rivera, N. & Kaminer, I. Light-matter interactions with photonic quasiparticles. *Nat. Rev. Phys.* **2**, 538-561 (2020).
[7] Hu, H., Lin, X., and Luo, Y. Free-electron radiation engineering via structured environments. *Progress in Electromagnetics Research* **171**, 75-88 (2021).
[8] Su, Z. et al. Manipulating Cherenkov radiation and Smith-Purcell radiation by artificial structures. *Adv. Opt. Mater.* **7**, 1801666 (2019).
[9] Wang, W. et al. Free-electron lasing at 27 nanometres based on a laser wakefield accelerator. *Nature* **595**, 516-520 (2021).
[10] Decking, W. et al. A MHz-repetition-rate hard X-ray free-electron laser driven by a superconducting linear accelerator. *Nat. Photon.* **14**, 391-397 (2020).
[11] McNeil, B. & Thompson, N. X-ray free-electron lasers. *Nat. Photon.* **4**, 814-821 (2010).
[12] Prat, E. et al. A compact and cost-effective hard X-ray free-electron laser driven by a high-brightness and low-energy electron beam. *Nat. Photon.* **14**, 748-754 (2020).
[13] Kaminsky, A. K. et al. Demonstrating high-power 30-GHz free-electron maser operation on a resonant load. *Tech. Phys. Lett.* **36**, 211-215 (2010).
[14] Gardelle, J., Labrouche, J. & Rullier, J. L. Direct observation of beam bunching produced by a high-power microwave free-electron laser. *Phys. Rev. Lett.* **76**, 4532-4535 (1996).
[15] Chamberlain, O., Segrè, E., Wiegand, C. & Ypsilantis, T. Observation of antiprotons. *Phys. Rev.* **100**, 947-950 (1955).
[16] Aubert, J. J. et al. Experimental observation of a heavy particle J. *Phys. Rev. Lett.* **33**, 1404-1406 (1974).
[17] Galbraith, W. & Jelley, J. V. Light pulses from the night sky associated with cosmic rays. *Nature* **171**, 349-350 (1953).
[18] Lin, X. et al. Controlling Cherenkov angles with resonance transition radiation. *Nat. Phys.* **14**, 816-821

(2018).
[19] Lin, X. et al. A Brewster route to Cherenkov detectors. *Nat. Commun.* **12**, 5554 (2021).
[20] Nozawa, I. et al. Measurement of <20 fs bunch length using coherent transition radiation. *Phys. Rev. ST Accel. Beams* **17**, 072803 (2014).
[21] Lumpkin, A. H. et al. First observation of z-dependent electron-beam microbunching using coherent transition radiation. *Phys. Rev. Lett.* **86**, 79 (2001).
[22] de Abajo, F. J. G. Optical excitations in electron microscopy. *Rev. Mod. Phys.* **82**, 209-275 (2010).
[23] Shaffer, T., Pratt, E. & Grimm, J. Utilizing the power of Cerenkov light with nanotechnology. *Nat. Nanotech.* **12**, 106-117 (2017).
[24] Dahan, R. et al. Creation of optical cat and GKP states using shaped free electrons. arXiv:2206.08828 (2022).
[25] Baranes, G. et al. Free electrons can induce entanglement between photons. *npj Quantum Inf.* **8**, 32 (2022).
[26] Huang, S. et al. Enhanced versatility of table-top X-rays from Van der Waals structures. *Adv. Sci.* **9**, 2105401 (2022).
[27] Huang, S. et al. Quantum recoil in free-electron interactions with atomic lattices. *Nat. Photon.* (2023). https://doi.org/10.1038/s41566-022-01132-6
[28] Wong, L. J. & Kaminer, I. Prospects in x-ray science emerging from quantum optics and nanomaterials. *Appl. Phys. Rev.* **119**, 130502 (2021).
[29] de Abajo, F. J. G. & Di Giulio, V. Optical excitations with electron beams: Challenges and opportunities. *ACS Photonics* **8**, 945-974 (2021).
[30] Roques-Carmes, C. et al. Free-electron-light interactions in nanophotonics. *Appl. Phys. Rev.* **10**, 011303 (2023).
[31] Sapra, N. V. et al. On-chip integrated laser-driven particle accelerator. *Science* **367**, 79-83 (2020).
[32] Madey, J. J. M. Stimulated emission of bremsstrahlung in a periodic magnetic field. *J. Appl. Phys.* **42**, 1906-1913 (1971).
[33] Cherenkov, P. A. Radiation from high-speed particles. *Science* **131**, 136-142 (1960).
[34] Frank, I. M. Optics of light sources moving in refractive media. *Science* **131**, 702-712 (1960).
[35] Shiloh, R. et al. Miniature light-driven nanophotonic electron acceleration and control. *Advances in Optics and Photonics* **14**, 862-932 (2022).
[36] Roques-Carmes, C. et al. Towards integrated tunable all-silicon free-electron light sources. *Nat Commun.* **10**, 3176 (2019).
[37] Hu, H. et al. Surface Dyakonov-Cherenkov radiation. *eLight* **2**, 2 (2022).
[38] Yang, Y. et al. Photonic flatband resonances for free-electron radiation. *Nature* **613**, 42-47 (2023).
[39] Ivanov, I. P. & Karlovets, D. V. Detecting transition radiation from a magnetic moment. *Phys. Rev. Lett.* **110**, 264801 (2013).
[40] Hu, H. et al. Broadband enhancement of Cherenkov radiation using dispersionless plasmons. *Adv. Sci.* **9**, 2200538 (2022).
[41] Xu, X. et al. Generation of terawatt attosecond pulses from relativistic transition radiation. *Phys. Rev. Lett.* **126**, 094801 (2021).
[42] Shentcis, M. et al. Tunable free-electron X-ray radiation from van der Waals materials. *Nat. Photon.* **14**, 686-692 (2020).
[43] Liu, F. et al. Integrated Cherenkov radiation emitter eliminating the electron velocity threshold. *Nat. Photon.* **11**, 289-292 (2017).
[44] Lin, X. et al. Splashing transients of 2D plasmons launched by swift electrons. *Sci. Adv.* **3**, e1601192 (2017).
[45] Duan, Z. et al. Observation of the reversed Cherenkov radiation. *Nat. Commun.* **8**, 14901 (2017).
[46] Wong, L. J., Kaminer, I., Ilic, O., Joannopoulos, J. D. & Soljačić, M. Towards graphene plasmon-based free-electron infrared to X-ray sources. *Nat. Photon.* **10**, 46-52 (2016).
[47] Genevet, P. et al. Controlled steering of Cherenkov surface plasmon wakes with a one-dimensional metamaterial. *Nat. Nanotech.* **10**, 804-809 (2015).
[48] Liu, S. et al. Surface polariton Cherenkov light radiation source. *Phys. Rev. Lett.* **109**, 153902 (2012).
[49] Tay, F. et al. Anomalous free-electron radiation beyond the conventional formation time. arXiv:2211.14377 (2022).
[50] Kolokolov, A. A. Reflection of plane-waves from an amplifying medium. *ZhETF Pis. Red.* **21**, 312-313 (1975).

[51] Lin, X. et al. Transverse-electric Brewster effect enabled by nonmagnetic two-dimensional materials. *Phys. Rev. A* **94**, 023836 (2016).
[52] Dorofeenko, A. V. et al. Light propagation in composite materials with gain layers. *Physics-Uspekhi* **55**, 1080 (2012).
[53] Wang, L. G., Wang, L., Al-Amri, M., Zhu, S.-Y. & Zubairy, M. S. Counterintuitive dispersion violating Kramers-Kronig relations in gain slabs. *Phys. Rev. Lett.* **112**, 233601 (2014).
[54] Kim, S. Y., and Vedam, K. Analytic solution of the pseudo-Brewster angle. *J. Opt. Soc. Am. A* **3**, 1772-1773 (1986).
[55] Lakhtakia, A. Would Brewster recognize today's Brewster angle? *Optics News* **15**, 14-18 (1989).
[56] Fan, H. et al. Brewster metasurfaces for ultrabroadband reflectionless absorption at grazing incidence. *Optica* **9**, 1138-1148 (2022).
[57] Luo, J. et al. Ultra-broadband reflectionless Brewster absorber protected by reciprocity. *Light: Science & Applications* **10**, 89 (2021).
[58] Kong, J. A. *Electromagnetic Wave Theory* (EMW Publishing, Cambridge Massachusetts, 2008).
[59] Ginzburg, V. L. Radiation of a uniformly moving electron due to its transition from one medium into another. *Sov. Phys. JETP* **16**, 15-28 (1946).
[60] Ginzburg, V. L. & Tsytovich, V. N. *Transition Radiation and Transition Scattering* (CRC Press, 1990).
[61] Chen, J. el al. Low-velocity-favored transition radiation. arXiv:2212.13066 (2022).
[62] Chen, R. et al. Recent advances of transition radiation: fundamentals and applications. *Materials Today Electronics* **3**, 100025 (2023).
[63] Garibyan, G. M. Transition radiation effects in particle energy losses. *Soviet Physics JETP* **37,** 527-533 (1959).
[64] Qian, C. et al. Breaking the fundamental scattering limit with gain metasurfaces. *Nat. Commun.* **13**, 4383 (2022).
[65] Ye, D., Chang, K., Ran, L. & Xin, H. Microwave gain medium with negative refractive index. *Nat. Commun.* **5**, 5841 (2014).
[66] Xu, W., Padilla, W. J., and Sonkusale, S. Loss compensation in metamaterials through embedding of active transistor based negative differential resistance circuits. *Opt. Express* **20,** 22406-22411 (2012).
[67] Jiang, T., Chang, K., Si, L.-M., Ran, L. & Xin, H. Active microwave negative-index metamaterial transmission line with gain. *Phys. Rev. Lett.* **107**, 205503 (2011).
[68] de Leon, I. & Berini, P. Amplification of long-range surface plasmons by a dipolar gain medium. *Nat. Photon.* **4**, 382-387 (2010).
[69] Gather, M. et al. Net optical gain in a plasmonic waveguide embedded in a fluorescent polymer. *Nat. Photon.* **4**, 457-461 (2010).
[70] Chen, J., Chen, H. & Lin, X. Photonic and plasmonic transition radiation from graphene. *Journal of Optics* **23**, 034001 (2021).
[71] Brewster, D. On the laws which regulate the polarization of light by reflection from transparent bodies. *Proceedings of the Royal Society of London Series I* **2**, 14-15 (1815).
[72] Paniagua-Domínguez, R. et al. Generalized Brewster effect in dielectric metasurfaces. *Nat. Commun.* **7**, 10362 (2016).
[73] Shen, Y. et al. Optical broadband angular selectivity. *Science* **343**, 1499-1501 (2014).

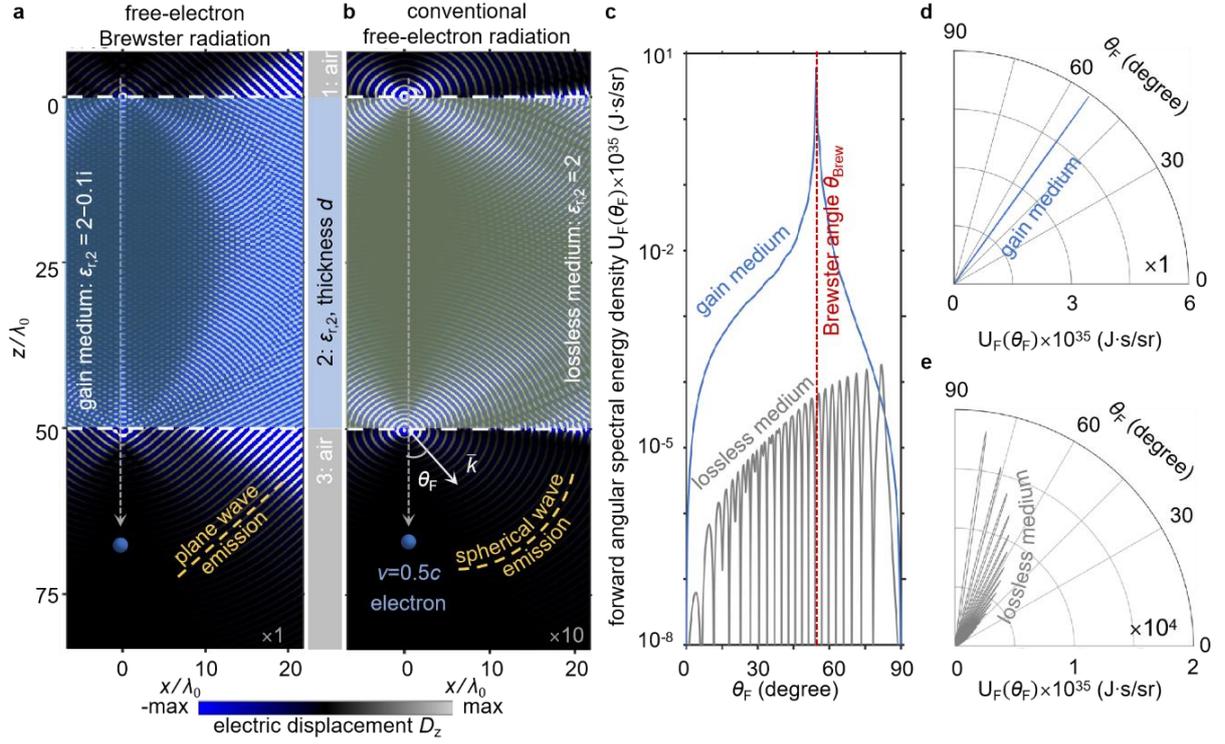

**Fig. 1 | Conceptual demonstration of free-electron Brewster radiation.** A moving electron perpendicularly penetrates through a slab with a relative permittivity of $\varepsilon_{r,2}$. **a, b**, Distribution of radiation field. The slab is constructed by a gain material with $-Im(\varepsilon_{r,2}) = 0.1i$ in (a) and a lossless material with $Im(\varepsilon_{r,2}) = 0$ in (b); see the structural schematic in the right side of (a). **c-e**, Angular spectral energy density $U_F(\theta_F)$ of forward free-electron radiation in the Cartesian and polar coordinates. $\theta_F$ is the angle between the wavevector of forward radiated light and the electron velocity.

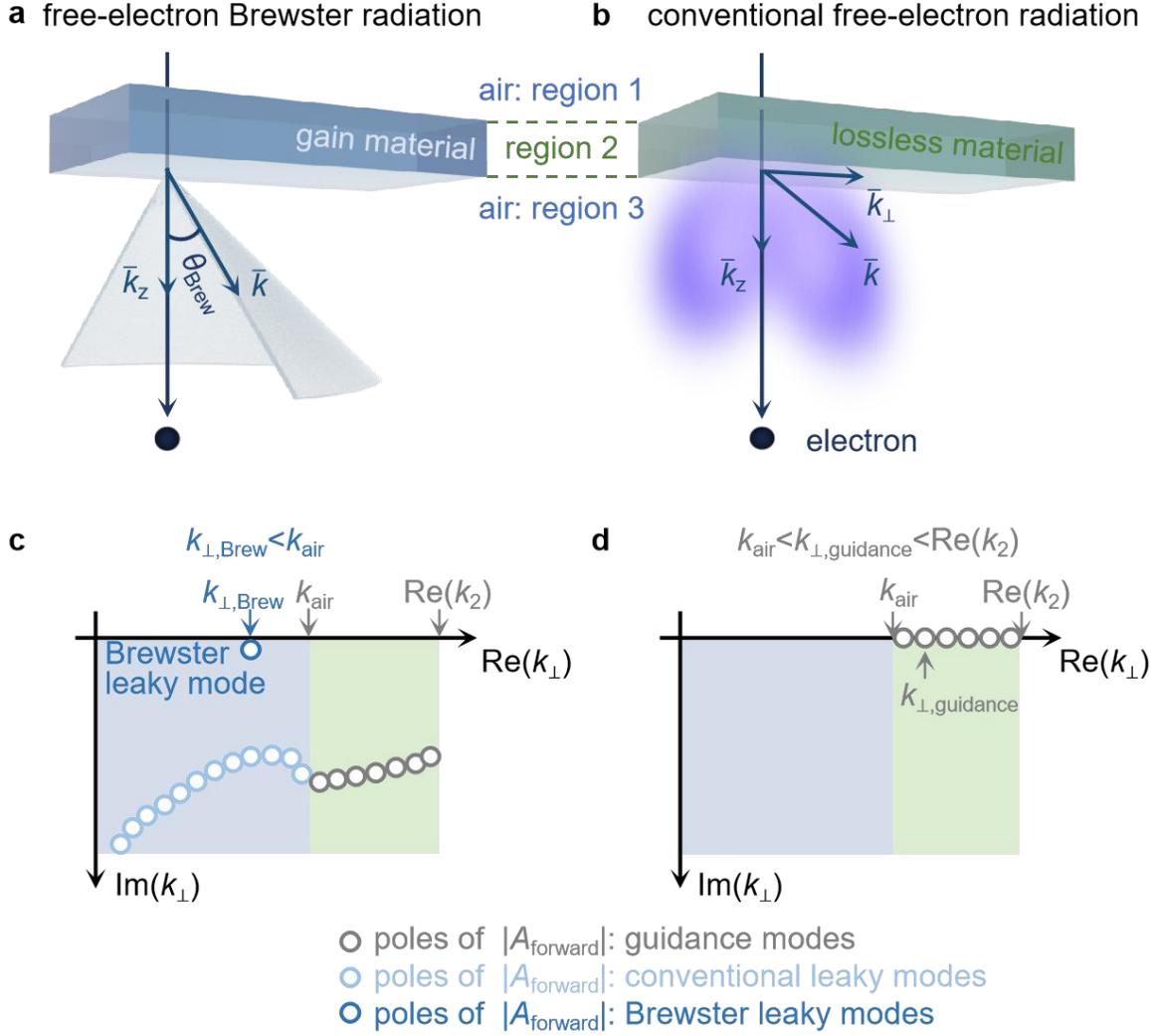

**Fig. 2 | Mechanism of free-electron Brewster radiation in *k*-space. a, b,** Schematic of free-electron radiation in *k*-space. The free-electron Brewster radiation in (a) is featured with an ultrahigh directionality at the radiation angle of $\theta_F = \theta_{Brew}$. By contrast, the conventional free-electron radiation in (b) is of low directionality. **c, d,** Poles or singularities of the forward radiation coefficient $|A_{forward}|$ in equation (2) in the complex $k_\perp$ plane. If region 2 is filled with a gain material with a sufficiently-large thickness in (c), there is one specific kind of pole always having $Re(k_\perp) = k_{\perp,Brew} = k_{air} sin\theta_{Brew}$. The corresponding mode is termed Brewster leaky mode, due to its close connection with the pseudo-Brewster effect of gain material.

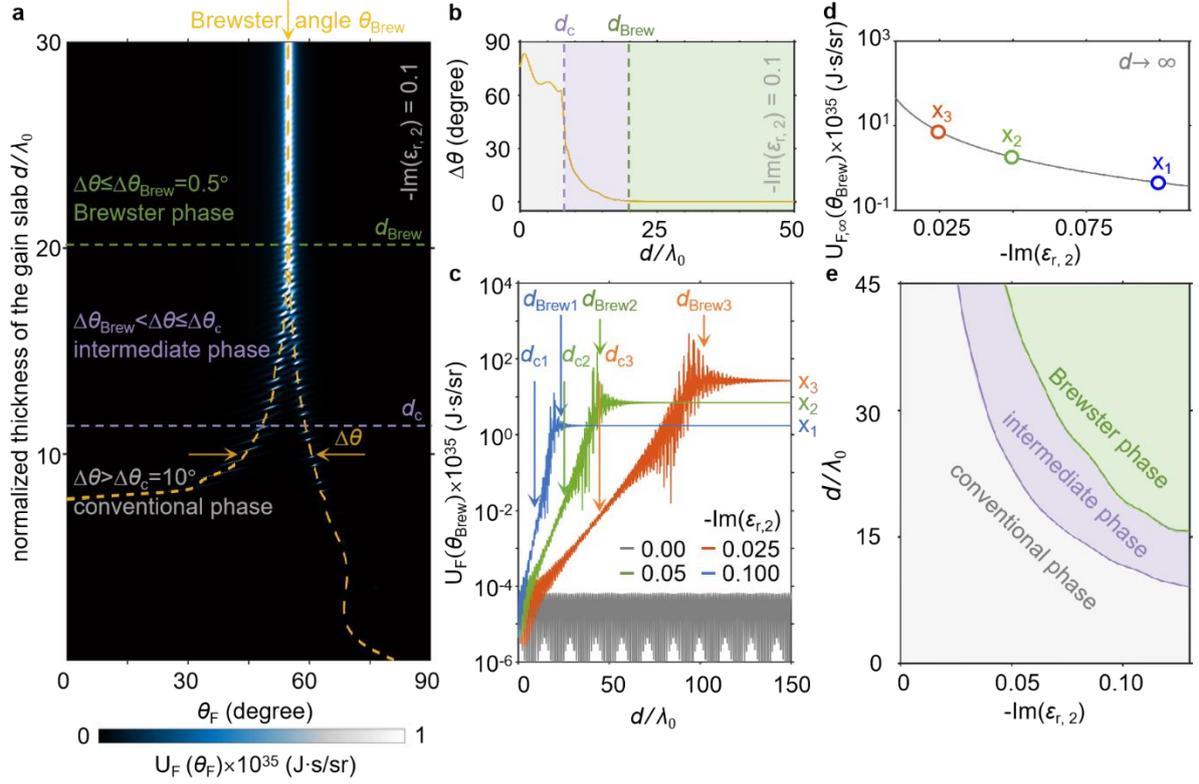

**Fig. 3 | Phase diagram and Brewster phase of free-electron radiation from a gain slab.** Here the structural setup is the same as Fig. 1, except for $-Im(\varepsilon_{r,2})$ and the slab thickness $d$. **a**, Angular spectral energy density $U_F(\theta_F)$ of forward free-electron radiation as a function of $\theta_F$ and $d$, with $-Im(\varepsilon_{r,2}) = 0.1$. The left and right yellow lines correspond to the trajectories of $\theta_{max,left} = \max(U_F(\theta_F))$ for $\theta_F \in [0°, \theta_{Brew}]$ and $\theta_{max,right} = \max(U_F(\theta_F))$ for $\theta_F \in [\theta_{Brew}, 90°])$, respectively. The angular deviation is defined as $\Delta\theta = \theta_{max,right} - \theta_{max,left}$. **b**, Dependence of $\Delta\theta$ on $d$, as extracted from (a). **c**, $U_F(\theta_{Brew})$ as a function of $d$ under different $-Im(\varepsilon_{r,2})$. If $-Im(\varepsilon_{r,2}) > 0$ and $d$ is large enough, the value of $U_F(\theta_{Brew})$ becomes a constant, which is denoted as $U_{F,\infty}(\theta_{Brew}) = \lim_{d\to\infty} U_F(\theta_{Brew})$. **d**, $U_{F,\infty}(\theta_{Brew})$ as a function of $-Im(\varepsilon_{r,2})$. **e**, Phase diagram of forward free-electron radiation in the parameter space of $-Im(\varepsilon_{r,2})$ and $d$, according to $\Delta\theta$. We have $\Delta\theta < \Delta\theta_{Brew}$ for the Brewster phase, $\Delta\theta_c > \Delta\theta > \Delta\theta_{Brew}$ for the intermediate phase, and $\Delta\theta > \Delta\theta_c$ for the conventional phase, where $\Delta\theta_{Brew} = 0.5°$ and $\Delta\theta_c = 10°$.

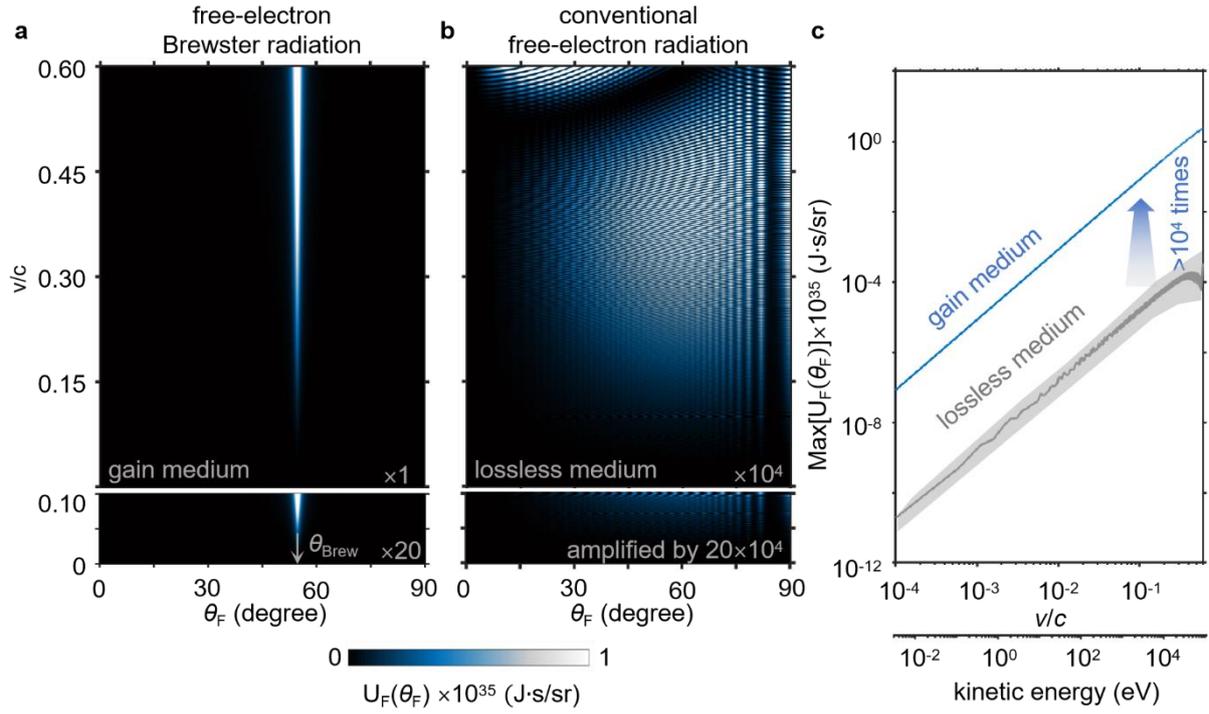

**Fig. 4 | Robustness of free-electron Brewster radiation with respect to the electron velocity.** Here the structural setup is the same as Fig. 1, except for the electron velocity $v$. **a**, **b**, Angular spectral energy density $U_F(\theta_F)$ of the forward free-electron radiation as a function of $v$ and $\theta_F$. The slab is filled with a gain medium in (a) and a lossless medium in (b). **c**, Dependence of $\max(U_F(\theta_F))$ on $v$, as extracted from (a, b).